\documentclass{emulateapj}
\usepackage{epsfig}
\usepackage{lscape} 
\submitted{{\it Accepted for Publication in PASP}} 
\newcommand{\simlt}%
        {\,\hbox{\lower0.6ex\hbox{$\sim$}\llap{\raise0.6ex\hbox{$<$}}}\,}
\shortauthors{Davenport et al.}
\shorttitle{Color Transformations for Cool Stars}
\begin{document}

\title{Sloan/Johnson-Cousins/2MASS Color Transformations for Cool-Stars}

\author{James R. A. Davenport}
\affil{Department of Astronomy, University of Washington, Box 351580,
Seattle, WA 98195}
\email{uwjim@astro.washington.edu}

\author{Andrew A. West}
\affil{Astronomy Department, University of California, 601 Campbell Hall, Berkeley, CA 94720-3411}
\email{awest@astro.berkeley.edu}

\author{Caleb K. Matthiesen}
\affil{Department of Astronomy, University of Washington, Box 351580,
Seattle, WA 98195}

\author{Michael Schmieding}
\affil{Department of Astronomy, University of Washington, Box 351580,
Seattle, WA 98195}

\author{Adam Kobelski}
\affil{Department of Astronomy, University of Washington, Box 351580,
Seattle, WA 98195}
\affil{Department of Physics, 264 EPS Building, Montana State University, Bozeman, MT 59717}

\begin{abstract} 
We present multi-color transformations and photometric parallaxes for
a sample of 40 low mass dwarfs selected from the Sloan Digital Sky
Survey (SDSS) and the General Catalog of Trigonometric Stellar
Parallaxes.  Our sample was re-observed at the Manastash Ridge
Observatory (MRO) using both Sloan and Johnson-Cousin filters and
color transformations between the two photometric systems were
derived.  A subset of the sample had previously measured
Johnson-Cousins photometry and parallaxes as well as 2MASS photometry.
We observed these stars at MRO using Sloan filters and used these data
to derive photometric parallax relations as well as
SDSS/Johnson-Cousins/2MASS color transformations.  We present the data
and derived transformations for use in future low mass star studies.
\end{abstract}


\section{Introduction}
The most numerous stars in the Galaxy are low-mass dwarfs and yet it
is only in the past couple of decades that large enough samples
existed to statistically study their properties.  The
Palomar/Michigan State University (PMSU) survey of low-mass stars
(Reid et al. 1995; Hawley et al. 1996; Gizis et al. 2002; Reid et
al. 2002;) examined the photometric, spectroscopic and dynamical
properties of low mass dwarfs with unprecedented statistical
accuracy. Recently, Bochanski et al. (2005) presented another
statistical study of the spectroscopic/kinematic properties of nearby
low mass dwarfs within 100 pc of the Sun.  The advent of large surveys
such as the Sloan Digital Sky Survey (SDSS; Gunn et al. 1998; Fukugita
et al. 1996; Lupton et al. 1999; York et al. 2000; Hogg et al. 2001;
Gunn et al. 2006; Ivezi{\'c} et al. 2004; Pier et al. 2003; Smith et
al. 2002; Stoughton et al. 2002) has increased the sample size and
statistical robustness of low-mass star studies (Hawley et al. 2002,
West et al. 2004; Silvestri et al. 2006; West et al. 2006; Bochanski
et al. 2006; Hawley et al. in prep).

One of the challenges with the SDSS samples has been the ability to
accurately compare the Sloan photometry to the Johnson-Cousins
standards for low-mass stars that are used in other studies.  Smith et
al. (2002) previously investigated the stellar SDSS/Johnson-Cousins
color transformation using the 158 SDSS standard stars.  However, the
SDSS standard stars only extend to an $r-i$ $<$ 1.5 (spectral type
$\sim$M4). Recently, Jordi, Grebel \& Ammomn (2006) and Rodgers
et al. (2006) examined a total of several hundred stars and
derived new SDSS/Johnson-Cousins transformations.  Unfortunately,
neither of these new sets of relations extend far beyond $r-i \sim1.5$ and
therefore are not useful for cool-star studies.

In another study, Walkowicz, Hawley \& West (2004) projected Sloan
filter curves onto spectra of stars with measured Johnson-Cousins
photometry in order to derive color transformations between the two
systems.  Although the Walkowicz et al. method has its utility and has
been used by several subsequent studies (West et al. 2004; West,
Walkowicz \& Hawley 2005; Silvestri et al. 2006), it is not the ideal
method for deriving color transformations.

Walkowicz et al. (2004) also derived a relation between Sloan colors
and absolute magnitude for stars with measured parallaxes.  However,
this relation used the intermediate step of estimating the Sloan
colors from either the spectral type of the star or the aforementioned
transformations.  No stars with known parallaxes were observed
directly with both filter sets.

In this paper, we undertake the more ideal solution of observing a set
of low-mass dwarfs with the same telescope in both Johnson-Cousins and
Sloan filters and using these data to derive a new set of color
transformations.  A subset of our stars have measured parallaxes and
2MASS photometry and we use these data to derive
SDSS/Johnson-Cousins/2MASS color transformations as well as new
photometric parallax relations for all three systems.  In \S2, we
describe our sample selection. We detail our observations in \S3 and
present our results in \S4. We discuss these our results and
possibilities for the future in \S5.

\section{Sample Selection}
\subsection{SDSS Selected Sample}
The SDSS/Johnson-Cousins color transformation sub-sample was selected
from the SDSS Data Release 4 (DR4; Abazajian et al. 2005) database.
We required that stars have colors 0.25 $< r-i <$ 3.5 and 0.75 $< g-r
<$ 2.0 based on the colors for cool-stars reported in West et
al. (2004, 2005) and Covey et al. (in prep).  In addition, we selected
stars with 15 $< r <$ 18.5, and to ensure a clean sample, made sure
that the SDSS flags SATURATED, EDGE, DEBLENDED\_AS\_MOVING, CHILD,
INTERP\_CENTER PSF\_FLUX\_INTERP and BLENDED were unset.  We
selected the 5 brightest, observable (based on RA) stars from every
0.1 bin in $r-i$ color (if 5 existed), resulting in a sample of 23
stars.  The sample was then divided into a ``blue'' sample ($r-i <$
1.4 ; 11 stars) and a ``red'' sample ($r-i >$ 1.4; 12 stars).

\subsection{Parallax Selected Sample}
The parallax sub-sample is composed of 17 stars selected from the
General Catalogue of Trigonometric Stellar Parallaxes (van Altena, Lee
\& Hoffleit 1995).  We first queried the catalog for those stars with
small parallax errors ($\sigma\simlt 0.01^{\prime\prime}$) and late spectral
types (K0 and later).  The final sub-sample was selected to span as
large a range of spectral types as possible, while maximizing
observational efficiency (including both location on the sky and
stellar magnitude).

2MASS photometry was obtained from the 2MASS point source catalog
(Skrutskie et al. 2006) and the Johnson-Cousin colors were taken from
the PMSU I (Reid et al. 1995).

\section{Observations}
Observations were carried out during the summer of 2005 using the
University of Washington 30-inch telescope at Manastash Ridge
Observatory (MRO).  We used the 2048 $\times$ 2048 CCD detector and
the MRO suite of $VRIriz$ filters.  Standard stars were observed to
derive extinction coefficients and photometric zero-points. Every
sample and standard star was observed long enough to obtain a minimum
of 10000 counts, and therefore a 1\% photon noise uncertainty.  The
MRO photometric camera was shown to be linear over the entire
magnitude range of our observations.  The data were reduced using the
standard IRAF packages.  All Sloan bands were reduced to the SDSS 2.5m
system (unprimed). The uncertainty in the derived extinction
coefficients dominates the reported photometric uncertainties.

For the SDSS selected sample, we observed each star using both the
Johnson-Cousins $RI$ filters and the Sloan $ri$ filters.  For the blue
sample, we added a Johnson-Cousins $V$ filter and for the red sample
the $z$ filter was included.  Because the stars in the SDSS selected
sample already had SDSS photometry, the MRO $riz$ data allow us ensure
that our photometry is correctly on the proper SDSS system.  We find
that the $r-i$ colors have a mean difference of 0.054 mags and an rms
scatter of 0.165 mags, while the $i-z$ colors have a mean offset of
-0.002 mags and an rms scatter of 0.220 mags.

\section{Results}
The resulting photometry can be found in Tables \ref{phot} and
\ref{phot2}.  The observational uncertainties reported in Tables
\ref{phot} and \ref{phot2} reflect the propagated uncertainties from
each stage of the photometric reductions. Using these new colors we
derived new color-color transformations between the Sloan,
Johnson-Cousins and 2MASS photometric systems.

\subsection{Color-Color Transformations}
We performed polynomial fits for each color-color relation and derive
the color transformation equations shown below.  Each equation
contains the uncertainty in the resulting transformed color.  The
equations and uncertainties derived in this section are only valid
over the range in which we have data.  These relations should not be
extrapolated to redder or bluer colors.

\begin{figure}
\includegraphics[scale=.5]{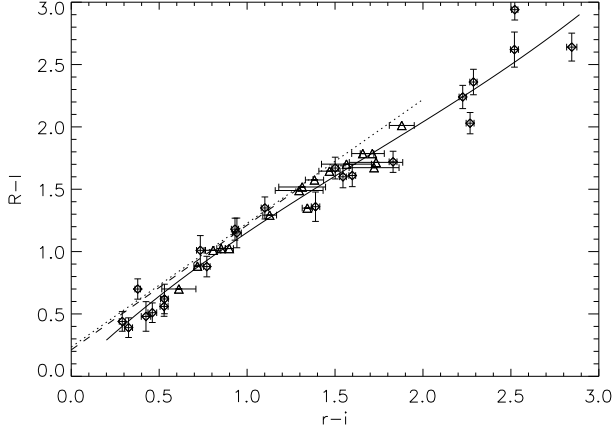}
\caption{$R-I$ vs. $r-i$ color transformation for low-mass dwarfs.
The derived polynomial relation is overplotted (solid) and is given by
Equation 1.  The relations from Jordi et al. (2006) and Rodgers et al. (2006) are overplotted for comparison (dotted and dashed respectively).  The three observationally derived relations agree to within the uncertainties.}
\label{RIri} 
\end{figure}

\begin{figure}
\includegraphics[scale=.5]{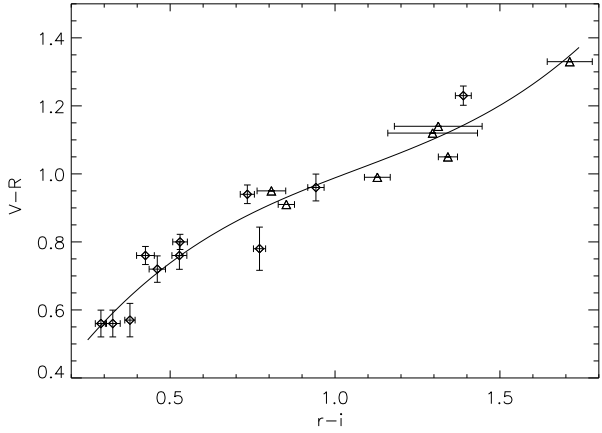}
\caption{$V-R$ vs. $r-i$ color transformation for the bluest stars in
our sample.  The derived polynomial relation is overplotted and is
given by Equation 2.}
\label{VRri} 
\end{figure}

The relations for $R-I$ and $V-R$ as a function of $r-i$ are shown in Figure \ref{RIri} and Figure \ref{VRri}, are given by:
 
\begin{eqnarray}
R - I & = & 0.14 + 0.38(r-i) + 0.29(r-i)^2\nonumber\\
 & & + 0.06(r-i)^3\;\pm0.12\\
\nonumber
\end{eqnarray}

\noindent and

\begin{eqnarray}
V - R & = & 0.17 + 1.63(r-i) - 1.19(r-i)^2\nonumber\\
 & &  + 0.37(r-i)^3\;\pm0.06\\
\nonumber
\end{eqnarray}

\noindent and are valid over the range $0.33 \le r-i \le$ 2.85. The $R-I$ vs. $r-i$ relations of Jordi et al. (2006) and Rodgers et al. (2006) have been overplotted (dotted and dashed respectively) on Figure \ref{RIri}.  The three relations agree within the uncertainties over the overlapping region.

\begin{figure}
\includegraphics[scale=.5]{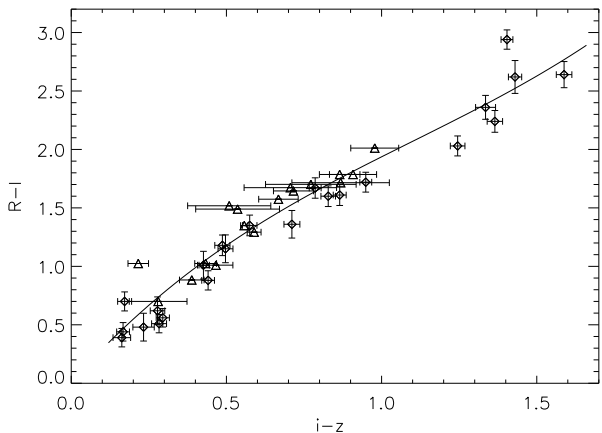}
\caption{$R-I$ vs. $i-z$ color transformation for low-mass dwarfs.
The derived polynomial relation is overplotted and is given by
Equation 3.} 
\label{RIiz} 
\end{figure}

The transformation between $R-I$ and $i-z$ is shown in Figure
\ref{RIiz}, described by:

\begin{eqnarray}
R - I & = & 0.01 + 2.98(i-z) - 1.51(i-z)^2\nonumber\\
& & + 0.46(i-z)^3\;\pm0.17\\
\nonumber
\end{eqnarray}

\noindent and is valid over the range 0.17 $\le i-z \le$ 1.59.

\begin{figure}
\includegraphics[scale=.5]{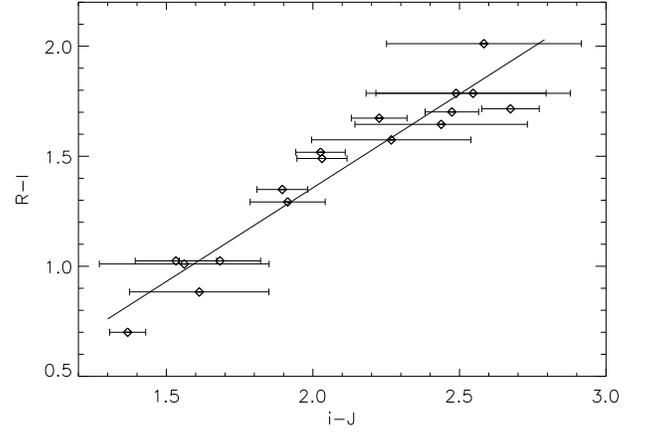}
\caption{$R-I$ vs. $i-J$ color transformation for stars with 2MASS
data.  The derived linear relation is overplotted and is given by
Equation 4.}
\label{RIij} 
\end{figure}

Although the uncertainties in $J$ are large, for completeness we included a first-order fit to the $R-I$ vs $i-J$ relation as Figure \ref{RIij} and given by:

\begin{equation}
R - I  =  -0.35 +0.85(i-J)\;\pm0.13
\end{equation}

\noindent This relation is valid over the range 1.37 $\le i-J \le$ 2.58.

We derived the color-color transformations for $r-i$ vs. $V-I$ to
compare with the same relation derived by West et al. (2005;
hereafter W05).  Although our sample does not go as red and is
slightly bluer than the W05. sample, the overlap region allows for
useful comparisons.  

\begin{figure}
\includegraphics[scale=.5]{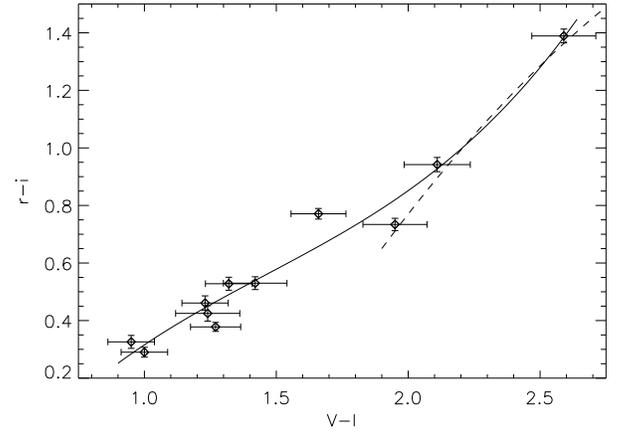}
\caption{$r-i$ vs. $V-I$ color transformation for the bluest stars in
our sample.  The derived polynomial relation is overplotted (solid)
and is given by Equation 5. The spectroscopically derived relation of
West et al. (2005) is also overplotted (dashed).}
\label{riVI} 
\end{figure}

Figure \ref{riVI} shows the $r-i$ vs. $V-I$ with the W05
transformation overplotted (dashed line).  Our new relation is
given by:

\begin{eqnarray}
r - i & = & -0.76 + 1.73(V-I) -0.85(V-I)^2\nonumber\\
& & + 0.19(V-I)^3\;\pm0.07\\
\nonumber
\end{eqnarray}
%
%

\noindent for the ranges 0.95 $\le V-I \le 2.59$.  Our relation agrees
with that of W05, confirming the accuracy and feasibility of the W05
method and extending the existing transformations to bluer colors.

\subsection{Photometric Parallax}

Using the stars with direct parallax measurements we derive
photometric parallax relations for the $i-z$ and $i-J$ colors.  These
relations are the same relations derived by W05.
Although some of the uncertainties in the stellar photometry are large,
these data serve as a direct confirmation of the W05 relations,
without having to transform to a different photometric system.

\begin{figure}
\includegraphics[scale=.5]{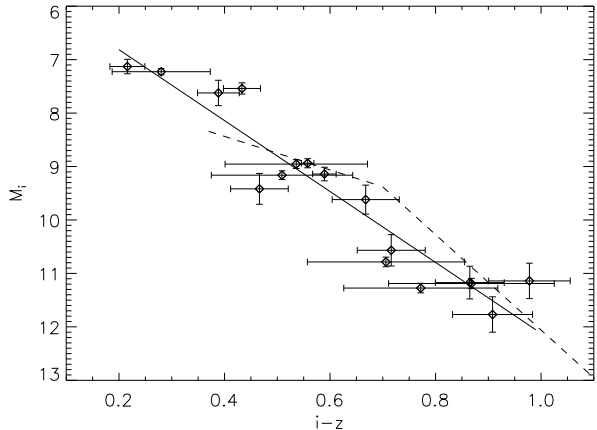}
\caption{Absolute $i$-band magnitude as a function of $i-z$ color for
the stars with direct trigonometric parallax measurements.  The
best-fit photometric parallax relation is plotted (solid) and given by
Equation 6. The West et al. (2005) relation is overplotted (dashed) for
comparison.}
\label{iiz} 
\end{figure}

\begin{figure}
\includegraphics[scale=.5]{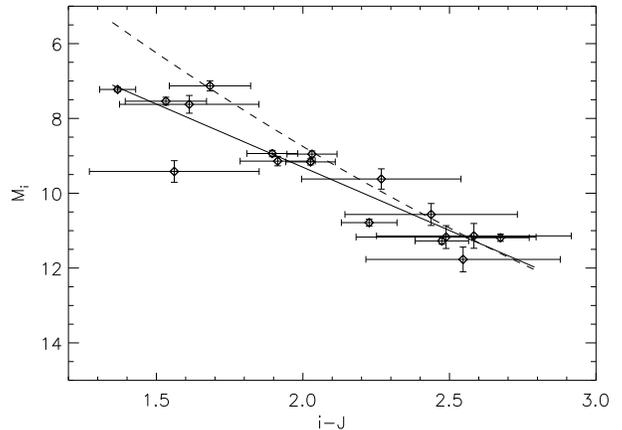}
\caption{Absolute $i$-band magnitude as a function of $i-J$ color for
the stars with direct trigonometric parallax measurements.  The
best-fit photometric parallax relation is plotted (solid) and given by
Equation 7. The West et al. (2005) relation is overplotted (dashed) for
comparison.} 
\label{iiJ} 
\end{figure}

The relations are shown in Figures \ref{iiz} and \ref{iiJ}.  The fits are plotted as solid lines and given by the equations,

\begin{equation}
M_{i}=5.49+6.63(i-z)\;\pm0.44\\
\end{equation}

\noindent and

\begin{equation}
M_{i}=2.55+3.38(i-J)\;\pm0.63\\
\end{equation}

and are valid over the ranges 0.22 $\le i-z \le$ 0.98 and 1.37 $\le
i-J \le$ 2.58 respectively.  The W05 photometric parallax relations
are shown as dashed lines.  Both of the fits from this study are
consistent with the W05 results and extend the relations to bluer
colors.  The $i-J$ relation of W05 can be extended to bluer colors and
appears to follow the same relation to within the uncertainties.
Although the piece-wise nature of the W05 $i-z$ relation precludes an
extrapolation to bluer colors, the $i-z$ relation from this paper
provides a useful extension for bluer stars.

\section{Discussion}

We have derived SDSS-Johnson-2MASS color transformations based on
direct measurements of cool-stars.  We present several relations not
included in previous studies of low-mass stars. Our $r-i$ vs. $V-I$
relation agrees with the previous findings of W05.  In addition, we
extend many of the relations to bluer colors than those included in
the W05 analysis and redder colors than the Smith et al. (2002),
Rodgers et al. (2006) and Jordi et al. (2006).  Table \ref{comp}
summarizes our derived color transformations and compares them with
the relations given in W05, Rodgers et al. (2006) and Jordi et
al. (2006).  This table is currently the most complete compilation of
low-mass star color transformations.

We also utilized the stars in our sample that have direct parallax
measurements to derive photometric parallax relations. Our results
confirm the relations given by W05.

All of our results indicate that the W05 method is justified and that
the spectroscopically derived W05 color transformations and
spectroscopic parallax relations are valid to within the quoted
uncertainties.  Future studies with larger samples of cool-stars will
further constrain these relations.

\acknowledgments The authors would like to thank Kevin Covey for his
useful discussions and many suggestions while working on this paper.
We would also like to thank Cindy Lee for helping with the
observations, Suzanne Hawley for her helpful suggestions and Chris
Laws for his support and care of the MRO facilities. AAW acknowledges
the support of NSF grant AST05-40567 and the financial support of
Julianne Dalcanton.

Funding for the SDSS and SDSS-II has been provided by the Alfred
P. Sloan Foundation, the Participating Institutions, the National
Science Foundation, the U.S. Department of Energy, the National
Aeronautics and Space Administration, the Japanese Monbukagakusho, the
Max Planck Society, and the Higher Education Funding Council for
England. The SDSS Web Site is http://www.sdss.org/.

The SDSS is managed by the Astrophysical Research Consortium for the
Participating Institutions. The Participating Institutions are the
American Museum of Natural History, Astrophysical Institute Potsdam,
University of Basel, Cambridge University, Case Western Reserve
University, University of Chicago, Drexel University, Fermilab, the
Institute for Advanced Study, the Japan Participation Group, Johns
Hopkins University, the Joint Institute for Nuclear Astrophysics, the
Kavli Institute for Particle Astrophysics and Cosmology, the Korean
Scientist Group, the Chinese Academy of Sciences (LAMOST), Los Alamos
National Laboratory, the Max-Planck-Institute for Astronomy (MPIA),
the Max-Planck-Institute for Astrophysics (MPA), New Mexico State
University, Ohio State University, University of Pittsburgh,
University of Portsmouth, Princeton University, the United States
Naval Observatory, and the University of Washington.

\clearpage

\begin{deluxetable}{ccrrrrrr}
\tablewidth{0pt}
\tablecolumns{8} 
\tabletypesize{\scriptsize}
\tablecaption{Photometry for SDSS Selected Stars}
\renewcommand{\arraystretch}{.6}
\tablehead{
\colhead{RA (J2000)}&
\colhead{DEC (J2000)}&
\colhead{$V-R$}&
\colhead{$R$}&
\colhead{$R-I$}&
\colhead{$r$}&
\colhead{$r-i$}&
\colhead{$i-z$}}
\startdata
12:34:57.5&$+$31:12:22&  0.94 (0.03)& 13.99 (0.71)&  1.01 (0.12)& 15.02 (0.02)&  0.73 (0.02)&  0.43 (0.02)\\
13:26:16.3&$+$56:40:44&\nodata \nodata& 13.44 (0.35)&  2.94 (0.08)& 17.68 (0.01)&  2.52 (0.02)&  1.40 (0.02)\\
13:49:06.4&$+$60:22:20&  0.76 (0.03)& 15.61 (0.71)&  0.48 (0.12)& 15.00 (0.02)&  0.43 (0.03)&  0.23 (0.03)\\
14:09:12.8&$+$65:00:00&\nodata \nodata& 15.49 (0.62)&  1.61 (0.09)& 15.91 (0.02)&  1.60 (0.02)&  0.86(0.02)\\ 
14:30:57.4&$+$51:24:52&  0.80 (0.02)& 15.38 (0.71)&  0.62 (0.12)& 15.01 (0.01)&  0.53 (0.02)&  0.28 (0.02)\\
14:44:17.1&$+$30:02:12&\nodata \nodata& 16.06 (0.62)&  2.64 (0.11)& 18.24 (0.02)&  2.85 (0.03)&  1.59 (0.03)\\
14:44:26.2&$+$62:01:25&  0.76 (0.04)& 15.35 (0.36)&  0.56 (0.08)& 15.00 (0.02)&  0.53 (0.02)&  0.30 (0.02)\\
15:06:24.5&$+$16:27:53&  0.72 (0.04)& 15.26 (0.36)&  0.51 (0.08)& 15.00 (0.02)&  0.46 (0.03)&  0.28 (0.02)\\
15:23:42.4&$+$35:47:09&\nodata \nodata& 14.30 (0.62)&  1.18 (0.09)& 15.06 (0.02)&  0.93 (0.02)&  0.49 (0.02)\\
15:27:07.4&$+$43:03:01&\nodata \nodata& 14.36 (0.62)&  1.35 (0.09)& 15.24 (0.01)&  1.10 (0.02)&  0.57 (0.02)\\
15:34:41.4&$+$46:41:32&\nodata \nodata& 17.06 (0.35)&  2.03 (0.09)& 17.43 (0.01)&  2.27 (0.02)&  1.24 (0.02)\\
15:46:05.3&$+$37:49:45&\nodata \nodata& 17.92 (0.37)&  2.62 (0.14)& 18.44 (0.02)&  2.52 (0.02)&  1.43 (0.02)\\
16:00:15.6&$+$52:52:27&  0.56 (0.04)& 15.39 (0.36)&  0.39 (0.08)& 15.00 (0.01)&  0.33 (0.02)&  0.16 (0.03)\\
16:11:39.8&$+$40:05:59&\nodata \nodata& 16.77 (0.35)&  1.72 (0.09)& 16.94 (0.02)&  1.83 (0.02)&  0.95 (0.02)\\
16:27:18.2&$+$35:38:35&\nodata \nodata& 15.93 (0.36)&  2.36 (0.10)& 17.72 (0.02)&  2.29 (0.02)&  1.34 (0.03)\\
16:37:18.6&$+$44:28:46&\nodata \nodata& 15.24 (0.61)&  1.60 (0.09)& 15.72 (0.01)&  1.55 (0.02)&  0.83 (0.02)\\
17:18:18.3&$+$65:12:12&  0.56 (0.04)& 15.34 (0.36)&  0.44 (0.08)& 15.00 (0.01)&  0.29 (0.02)&  0.17 (0.02)\\
17:30:17.3&$+$62:29:26&  1.23 (0.03)& 15.80 (0.71)&  1.36 (0.12)& 15.45 (0.02)&  1.39 (0.02)&  0.71 (0.03)\\
21:36:32.5&$-$06:41:51&  0.57 (0.05)& 13.89 (0.36)&  0.70 (0.08)& 15.02 (0.01)&  0.38 (0.02)&  0.17 (0.02)\\
21:37:27.5&$-$08:27:13&  0.78 (0.06)& 13.86 (0.36)&  0.88 (0.08)& 15.02 (0.01)&  0.77 (0.02)&  0.44 (0.02)\\
23:21:24.0&$-$09:57:36&  0.96 (0.04)& 14.32 (0.71)&  1.15 (0.12)& 15.01 (0.02)&  0.94 (0.03)&  0.50 (0.02)\\
23:46:44.7&$+$16:03:59&\nodata \nodata& 16.47 (0.62)&  2.24 (0.09)& 17.38 (0.01)&  2.23 (0.02)&  1.36 (0.03)\\
23:50:47.8&$+$14:42:44&\nodata \nodata& 15.21 (0.35)&  1.67 (0.09)& 16.39 (0.01)&  1.50 (0.02)&  0.79 (0.02)\\
\enddata
\tablecomments{The Sloan photometry comes from SDSS DR4 database. The Johnson/Cousins magnitudes were measured from MRO data.  Photometric uncertainties are included in parentheses.}
\label{phot}
\end{deluxetable}

\vskip 1in

\begin{deluxetable*}{ccrrrrrrrrrrrr}
\tablewidth{0pt}
\tablecolumns{11} 
\tabletypesize{\tiny}
\tablecaption{Photometry for Parallax Selected Sample}
\renewcommand{\arraystretch}{.6}
\tablehead{
\colhead{}&
\colhead{}&
\multicolumn{5}{c}{MRO\tablenotemark{a}~/2MASS\tablenotemark{b}}&
\colhead{parallax\tablenotemark{c}}&
\multicolumn{3}{c}{PMSU\tablenotemark{d}}\\
\cline{3-7}\cline{9-11}\\
\colhead{RA (J2000)}&
\colhead{DEC (J2000)}&
\colhead{$r-i$}&
\colhead{$i$}&
\colhead{$i-z$}&
\colhead{$J$}&
\colhead{$i-J$}&
\colhead{($^{\prime\prime}$)}&
\colhead{$V$}&
\colhead{$R$}&
\colhead{$I$}}
\startdata
      12:49:02.7& $+$66:06:37 & 1.30 (0.14)&  8.91 (0.08)&  0.54 (0.14)&  6.88 (0.02)&  2.03 (0.09)&0.1020 (0.0047)& 10.87&  9.75&  8.28\\
      13:19:45.7& $+$47:46:39 &  0.85 (0.03)&  7.02 (0.14)&  0.22 (0.03)&  5.34 (0.03)&  1.68 (0.14)&0.1050 (0.0061)&  9.07&  8.16&  7.18\\
      14:02:33.3& $+$46:20:24 &  0.90 (0.02)&  7.79 (0.11)&  0.43 (0.04)&  6.26 (0.09)&  1.53 (0.14)&0.0890 (0.0039)&\nodata&\nodata&\nodata\\
      14:53:51.2& $+$23:33:21 & 1.38 (0.05)&  9.71 (0.27)&  0.67 (0.06)&  7.44 (0.02)&  2.27 (0.27)&0.0960 (0.0041)&\nodata&\nodata&\nodata\\
      16:02:50.6& $+$20:35:16 & 1.47 (0.06)& 10.57 (0.29)&  0.72 (0.06)&  8.13 (0.02)&  2.44 (0.29)&0.1000 (0.0031)&\nodata&\nodata&\nodata\\
      16:24:09.5& $+$48:21:10 & 1.13 (0.04)&  8.55 (0.13)&  0.59 (0.02)&  6.64 (0.02)&  1.91 (0.13)&0.1310 (0.0048)& 10.73&  9.74&  8.43\\
      17:07:07.4& $+$21:33:14 & 1.34 (0.03)&  9.78 (0.09)&  0.56 (0.01)&  7.88 (0.02)&  1.90 (0.09)&0.0680 (0.0029)& 10.96&  9.91&  8.55\\
      17:19:54.2& $+$26:30:04 & 1.31 (0.13)&  9.30 (0.08)&  0.51 (0.13)&  7.27 (0.02)&  2.03 (0.08)&0.0940 (0.0021)& 11.03&  9.89&  8.36\\
      18:35:20.3& $+$45:45:06 &  0.72 (0.02)&  8.49 (0.24)&  0.39 (0.04)&  6.88 (0.02)&  1.61 (0.24)&0.0670 (0.0020)&\nodata&\nodata&\nodata\\
      19:21:34.0& $+$20:50:24 & 1.57 (0.14)& 11.27 (0.09)&  0.77 (0.15)&  8.80 (0.02)&  2.47 (0.09)&0.1000 (0.0030)&\nodata&\nodata&\nodata\\
      20:03:24.8& $+$29:52:00 & 1.71 (0.07)& 12.10 (0.33)&  0.91 (0.06)&  9.55 (0.02)&  2.55 (0.33)&0.0860 (0.0228)& 13.34& 12.01& 10.23\\
      20:05:02.8& $+$54:25:43 & 0.81 (0.04)& 10.49 (0.29)&  0.47 (0.06)&  8.93 (0.02)&  1.56 (0.29)&0.0610 (0.0022)& 10.98& 10.03&  8.99\\
      20:29:53.1& $+$09:40:17 & 1.73 (0.15)& 10.90 (0.10)&  0.87 (0.16)&  8.23 (0.02)&  2.67 (0.10)&0.1140 (0.0019)&\nodata&\nodata&\nodata\\
      20:40:40.4& $+$15:29:51 & 1.66 (0.06)& 11.13 (0.31)&  0.86 (0.07)&  8.64 (0.03)&  2.49 (0.31)&0.1020 (0.0102)&\nodata&\nodata&\nodata\\
      20:41:28.4& $+$57:25:08 & 0.61 (0.10)&  9.01 (0.06)&  0.28 (0.09)&  7.64 (0.02)&  1.37 (0.06)&0.0440 (0.0102)&\nodata&\nodata&\nodata\\
      20:43:25.8& $+$55:22:55 & 1.88 (0.07)& 12.14 (0.33)&  0.98 (0.08)&  9.56 (0.02)&  2.58 (0.33)&0.0630 (0.0055)&\nodata&\nodata&\nodata\\
      23:43:11.8& $+$36:32:08 & 1.72 (0.14)& 10.34 (0.09)&  0.71 (0.15)&  8.11 (0.03)&  2.23 (0.10)&0.1230 (0.0029)&\nodata&\nodata&\nodata\\
\enddata
\tablenotetext{a}{Sloan magnitudes and colors derived from MRO photometric data.}
\tablenotetext{b}{2MASS point source catalog photometry (Skrutskie et al. 2006).}
\tablenotetext{c}{Parallaxes from General Catalogue of Trigonometric Stellar Parallaxes (van Altena et al. 1995).}
\tablenotetext{d}{Johnson-Cousins photometry from PMSU I (Reid et al. 1995).}
\tablecomments{Uncertainties are included in parentheses where available.}
\label{phot2}
\end{deluxetable*}

\clearpage
    
\begin{landscape}
\begin{deluxetable*}{cccc}
\tablewidth{0pt}
\tablecolumns{4}
\tabletypesize{\tiny}
\tablecaption{Summary of Color Transformations}
\tablehead{
\colhead{Davenport et al.\tablenotemark{a}}&
\colhead{West et al. (2005)\tablenotemark{b}}&
\colhead{Rodgers et al. (2006)\tablenotemark{c}}&
\colhead{Jordi et al. (2006)\tablenotemark{d}}}
\startdata
$V - R  =  0.17 + 1.63(r-i) - 1.19(r-i)^2$&\nodata&\nodata&$g - r = (1.646\;\pm0.008)(V - R)
$\\
 $+0.37(r-i)^3\;\pm0.06$& & &$- 0.139\;\pm0.004$\\

$R - I  =  0.14 + 0.38(r-i)+ 0.29(r-i)^2$&\nodata&$r - i = (1.000\;\pm0.006)(R - I) - 0.212$&$r - i = (1.007\;\pm0.005)(R - I)$\\

$+0.06(r-i)^3\;\pm0.12$& & &$- 0.236\;\pm0.003$\\

$R - I  =  0.01 + 2.98(i-z)- 1.51(i-z)^2$&\nodata&$r - z = (1.567\;\pm0.020)(R - I) -0.365$&$r - z = (1.584\;\pm0.008)(R - I)$\\

$+0.46(i-z)^3\;\pm0.17$& & &$- 0.386\;\pm0.005$\\

$R - I  =  -0.35 +0.85(i-J)\;\pm0.13$ &$i - z = -20.6 +26.0(I - K) -
11.7(I - K)^2$&\nodata&\nodata\\

 &$+ 2.30(I - K)^3 - 0.17(I - K)^4$& & \\

$r - i = -0.76 + 1.73(V-I)-0.85(V-I)^2$&$r - i = -2.69 + 2.29(V - I) - 0.28(V - I)^2$&\nodata&\nodata\\

$+ 0.19(V-I)^3$ & & & \\
\enddata
\tablenotetext{a}{Relations are valid over 0.17 $\le i-z \le$ 1.59, $0.33 \le r-i \le$ 2.85, 1.37 $\le i-J \le$ 2.58, and 0.95 $\le V-I \le$ 2.59}
\tablenotetext{b}{Relations are valid over 0.67 $\le r-i \le$ 2.01, and 0.37 $\le i-z \le$ 1.84}
\tablenotetext{c}{Relations are valid over -0.095 $\le R-I \le$ 0.695}
\tablenotetext{d}{Relations are valid over 0.0 $\le V-R \le$ 1.2, and 0.0 $\le R-I \le$ 2.0}
\label{comp}
\end{deluxetable*}

\clearpage
\end{landscape}


\begin{thebibliography}{}

\bibitem[{{Abazajian} {et~al.}(2004)}]{2004AJ....128..502A}
{Abazajian}, K., {et~al.} 2004, \aj, 128, 502

\bibitem[{{Adelman-McCarthy} {et~al.}(2006)}]{2006ApJS..162...38A}
{Adelman-McCarthy}, J.~K., {et~al.} 2006, \apjs, 162, 38

\bibitem[{{Bochanski} {et~al.}(2005){Bochanski}, {Hawley}, {Reid}, {Covey},
  {West}, {Tinney}, \& {Gizis}}]{2005AJ....130.1871B}
{Bochanski}, J.~J., {Hawley}, S.~L., {Reid}, I.~N., {Covey}, K.~R., {West},
  A.~A., {Tinney}, C.~G., \& {Gizis}, J.~E. 2005, \aj, 130, 1871

\bibitem[{{Bochanski} {et~al.}(2006){Bochanski}, {West}, {Hawley}, \&
  {Covey}}]{2003AJ}
{Bochanski}, J.~J., {West}, A.~A., {Hawley}, S.~L., \& {Covey}, K.~R. 2006,
  \aj, in press

\bibitem[{{Fukugita} {et~al.}(1996){Fukugita}, {Ichikawa}, {Gunn}, {Doi},
  {Shimasaku}, \& {Schneider}}]{1996AJ....111.1748F}
{Fukugita}, M., {Ichikawa}, T., {Gunn}, J.~E., {Doi}, M., {Shimasaku}, K., \&
  {Schneider}, D.~P. 1996, \aj, 111, 1748

\bibitem[{{Gizis} {et~al.}(2002){Gizis}, {Reid}, \&
  {Hawley}}]{2002AJ....123.3356G}
{Gizis}, J.~E., {Reid}, I.~N., \& {Hawley}, S.~L. 2002, \aj, 123, 3356

\bibitem[{{Gunn} {et~al.}(1998)}]{1998AJ....116.3040G}
{Gunn}, J.~E., {et~al.} 1998, \aj, 116, 3040

\bibitem[{{Gunn} {et~al.}(2006)}]{2006AJ....131.2332G}
---. 2006, \aj, 131, 2332

\bibitem[{{Hawley} {et~al.}(2002)}]{2002AJ....123.3409H}
{Hawley}, S.~L., {Covey}, K.~R., {et~al.} 2002, \aj, 123, 3409

\bibitem[{{Hawley} {et~al.}(1996){Hawley}, {Gizis}, \&
  {Reid}}]{1996AJ....112.2799H}
{Hawley}, S.~L., {Gizis}, J.~E., \& {Reid}, I.~N. 1996, \aj, 112, 2799

\bibitem[{{Hogg} {et~al.}(2001){Hogg}, {Finkbeiner}, {Schlegel}, \&
  {Gunn}}]{2001AJ....122.2129H}
{Hogg}, D.~W., {Finkbeiner}, D.~P., {Schlegel}, D.~J., \& {Gunn}, J.~E. 2001,
  \aj, 122, 2129

\bibitem[{{Ivezi{\'c}} {et~al.}(2004)}]{2004AN....325..583I}
{Ivezi{\'c}}, {\v Z}., {et~al.} 2004, Astronomische Nachrichten, 325, 583

\bibitem[{{Jordi} {et~al.}(2006)}]{2006AA....}
{Jordi}, K., {Grebel} E.~K., \& {Ammon}, K. 2006, \aap, in press

\bibitem[{{Lupton} {et~al.}(1999){Lupton}, {Gunn}, \&
  {Szalay}}]{1999AJ....118.1406L}
{Lupton}, R.~H., {Gunn}, J.~E., \& {Szalay}, A.~S. 1999, \aj, 118, 1406

\bibitem[{{Pier} {et~al.}(2003){Pier}, {Munn}, {Hindsley}, {Hennessy}, {Kent},
  {Lupton}, \& {Ivezi{\' c}}}]{2003AJ....125.1559P}
{Pier}, J.~R., {Munn}, J.~A., {Hindsley}, R.~B., {Hennessy}, G.~S., {Kent},
  S.~M., {Lupton}, R.~H., \& {Ivezi{\' c}}, {\v Z}. 2003, \aj, 125, 1559

\bibitem[{{Reid} {et~al.}(2002){Reid}, {Gizis}, \&
  {Hawley}}]{2002AJ....124.2721R}
{Reid}, I.~N., {Gizis}, J.~E., \& {Hawley}, S.~L. 2002, \aj, 124, 2721

\bibitem[{{Reid} {et~al.}(1995){Reid}, {Hawley}, \&
  {Gizis}}]{1995AJ....110.1838R}
{Reid}, I.~N., {Hawley}, S.~L., \& {Gizis}, J.~E. 1995, \aj, 110, 1838

\bibitem[Rodgers et al.(2006)]{2006AJ....132..989R} Rodgers, C.~T., 
Canterna, R., Smith, J.~A., Pierce, M.~J., \& Tucker, D.~L.\ 2006, \aj, 
132, 989 

\bibitem[{{Silvestri} {et~al.}(2006)}]{2006AJ....131.1674S}
{Silvestri}, N.~M., {et~al.} 2006, \aj, 131, 1674

\bibitem[{{Skrutskie} {et~al.}(2006)}]{2006AJ....131.1163S}
{Skrutskie}, M.~F., {et~al.} 2006, \aj, 131, 1163

\bibitem[{{Smith} {et~al.}(2002)}]{2002AJ....123.2121S}
{Smith}, J.~A., {et~al.} 2002, \aj, 123, 2121

\bibitem[{{Stoughton} {et~al.}(2002)}]{2002AJ....123..485S}
{Stoughton}, C., {et~al.} 2002, \aj, 123, 485

\bibitem[{{Strauss} {et~al.}(2002)}]{2002AJ....124.1810S}
{Strauss}, M.~A., {et~al.} 2002, \aj, 124, 1810

\bibitem[{{van Altena} {et~al.}(1995){van Altena}, {Lee}, \&
  {Hoffleit}}]{1995gcts.book.....V}
{van Altena}, W.~F., {Lee}, J.~T., \& {Hoffleit}, E.~D. 1995, {The general
  catalogue of trigonometric [stellar] paralaxes} (New Haven, CT: Yale
  University Observatory, |c1995, 4th ed., completely revised and enlarged)

\bibitem[{{Walkowicz} {et~al.}(2004){Walkowicz}, {Hawley}, \&
  {West}}]{2004PASP..116.1105W}
{Walkowicz}, L.~M., {Hawley}, S.~L., \& {West}, A.~A. 2004, \pasp, 116, 1105

\bibitem[{{West} {et~al.}(2004){West}, {Hawley}, {Walkowicz}, {Covey},
  {Silvestri}, {Raymond}, {Harris}, {Munn}, {McGehee}, {Ivezi{\' c}}, \&
  {Brinkmann}}]{2004AJ....128..426W}
{West}, A.~A., {Hawley}, S.~L., {Walkowicz}, L.~M., {Covey}, K.~R.,
  {Silvestri}, N.~M., {Raymond}, S.~N., {Harris}, H.~C., {Munn}, J.~A.,
  {McGehee}, P.~M., {Ivezi{\' c}}, {\v Z}., \& {Brinkmann}, J. 2004, \aj, 128,
  426

\bibitem[{{West} {et~al.}(2005){West}, {Walkowicz}, \&
  {Hawley}}]{2005PASP..117..706W}
{West}, A.~A., {Walkowicz}, L.~M., \& {Hawley}, S.~L. 2005, \pasp, 117, 706

\bibitem[{{West} {et~al.}(2006)}]{2006AJ.....}
{West}, A.~A., {Bochanski}, J.~J., {Hawley}, S.~L., {Cruz}, K.~L., {Covey}, K.~R., {Silvestri}, N.~M., {Reid}, I.~N., \& {Liebert}, J. 2006, \aj, in press


\bibitem[{{York} {et~al.}(2000)}]{2000AJ....120.1579Y}
{York}, D.~G., {et~al.} 2000, \aj, 120, 1579

\end{thebibliography}
\end{document}